# Analogies between Jahn-Teller and Rashba spin physics

Alessandro Stroppa[1,†,*], Paolo Barone[1,†], Domenico Di Sante[1,2], Mario Cuoco[3], Silvia Picozzi[1] and Myung-Hwan Whangbo[4,*]

[1] Consiglio Nazionale delle Ricerche (CNR-SPIN), I-67100 L'Aquila, Italy

[2] Department of Physical and Chemical Sciences, University of L'Aquila, I-67100 L'Aquila, Italy.

[3] Consiglio Nazionale delle Ricerche (CNR-SPIN) and Dipartimento di Fisica "E. R. Caianiello", Universitá degli Studi di Salerno, I-84084 Fisciano (SA), Italy

[4] Department of Chemistry, North Carolina State University, Raleigh, North Carolina 27695-8204, USA

† These authors contributed equally.

* Corresponding authors:      mike_whangbo@ncsu.edu

                                            alessandro.stroppa@spin.cnr.it

**Abstract**

In developing physical theories analogical reasoning has been found to be very powerful, as attested by a number of important historical examples. An analogy between two apparently different phenomena, once established, allows one to transfer information and bring new concepts from one phenomenon to the other. Here we discuss an important analogy between two widely different physical problems, namely, the Jahn-Teller distortion in molecular physics and the Rashba spin splitting in condensed matter physics. By exploring their conceptual and mathematical features and by searching for the counterparts between them, we examine the orbital texture in Jahn-Teller systems, as the counterpart of the spin texture of the Rashba physics, and put forward a possible way of experimentally detecting the orbital texture. Finally, we discuss the analogy by comparing the coexistence of linear Rashba+Dresselhaus effects and Jahn-Teller problems for specific symmetries, which allow for non-trivial spin and orbital textures, respectively.

**Introduction**

Reasoning in terms of analogies has been a powerful tool for developing new theories in mathematics and physics. Important historical examples in physics include Maxwell's application of fluid mechanics to formulate the theory of electromagnetism[1] and de Broglie's postulation of matter waves from the wave-particle duality of light quanta,[2] to name a few. Recently, it has been recognized that photonics is useful for mimicking other physical systems observed in astrophysics and hydrodynamics.[3] A formal analogy, often represented mathematically, occurs when the same relations hold for different systems despite no material similarity between them.[4] Analogies between different physical problems allow one to transfer information and bring new concepts from one phenomenon to the other. Analogies on conceptual level are also important in understanding the same physics behind two seemingly unrelated phenomena, as found for Jahn-Teller (JT) distortion[5] in discrete molecules and charge density wave formation in low-dimensional metals.[6]

In this work we show an important analogy between widely different problems, namely, the JT distortion[5] in molecular physics and the Rashba spin splitting[7-9] in condensed matter physics. Although the two phenomena differ in their physical context, we establish a formal mathematical mapping between them, explore their conceptual features, and search for the counterparts between them. As a natural extension, we define in the JT theory the orbital texture, as the counterpart of the spin texture in spin physics, and propose a possible way of detecting it experimentally. Furthermore, we discuss the analogy between related spin-splitting phenomena (such as linear Dresselhaus effect, possibly coexisting with the Rashba effect) and JT problems for specific symmetry, which lead to complex spin and orbital polarizations.

**Jahn-Teller distortion**

We begin our discussion with the JT distortion. When two potential surfaces of a given system are degenerate (or nearly degenerate) at some point $R_0$ in the nuclear coordinate space $R$, the topology of the energy surfaces, $E(R)$ vs. $R$, near the intersection point $R_0$ are classified into two groups depending whether or not the gradients $\nabla E(R)$ vanishes as $R \rightarrow R_0$. If the gradients are nonzero, then the surfaces around $R_0$ have the topology of a "conical intersection", leading to a JT system. If the gradients go to zero, then $R_0$ is a stationary point, leading to a Renner-Teller system. A JT system is subject to the JT theorem that any electronically degenerate system is intrinsically unstable under a certain symmetry-lowering distortion of the nuclear framework leading to the lifting of the degeneracy. In the following we discuss one of the simplest and most widespread JT problem, the E⊗ε problem, which applies to a variety of systems and displays a doublet of electronically degenerate states (E) interacting linearly with a doublet of degenerate displacements (ε). Consider the electronic doublet composed of two functions $\phi_x$ and $\phi_y$. Using the doubly degenerate vibrational modes, $q_1$ and $q_2$, the electron-lattice coupling matrix up to linear order can be written as[10]

$$V(q_1, q_2) = \gamma \begin{pmatrix} -q_1 & q_2 \\ q_2 & q_1 \end{pmatrix} \tag{1}$$

where the constant $\gamma$ represents the strength of the electron-phonon coupling that depends on a specific system, while the matrix structure is universal for any E⊗ε JT problem. Eq. 1 shows that $q_1$ splits the degeneracy of the electronic states ($\phi_x$, $\phi_y$) while $q_2$ introduces a mixing between the two. If the kinetic energy of the nuclei are neglected, the coupled vibronic system E⊗ε is described by the JT Hamiltonian, $H_{JT}$

$$H_{JT} = \gamma \begin{pmatrix} -q_1 & q_2 \\ q_2 & q_1 \end{pmatrix} + \frac{1}{2} C \left( q_1^2 + q_2^2 \right) \begin{pmatrix} 1 & 0 \\ 0 & 1 \end{pmatrix} \tag{2}$$

After introducing the polar coordinates $(q, \theta)$ in the $(q_1, q_2)$ plane with $q = \sqrt{q_1^2 + q_2^2}$, $q_1 = q\cos\theta$, and $q_2 = q\sin\theta$, the diagonalization of the 2×2 matrix leads to the eigenvalues[12]

$$E_{\mp}(q) = \mp \gamma q + \frac{1}{2} C q^2 \tag{3}$$

where γ is assumed to be positive without loss of generality. Then the minimum of the lower-energy state $E_-(q)$ occurs at $q = q_0 = \gamma/C$, leading to the energy lowering $E_0 = E_-(q) = -\gamma^2/2C$. (Here the quantity $|E_0|$ is known as the JT energy, $E_{JT}$.) One can express the associated eigenfunctions as

$$\left|\Psi_-\right\rangle = e^{-i\theta/2}\cos(\theta/2)\left|\phi_x\right\rangle - e^{-i\theta/2}\sin(\theta/2)\left|\phi_y\right\rangle$$
$$\left|\Psi_+\right\rangle = e^{+i\theta/2}\cos(\theta/2)\left|\phi_x\right\rangle + e^{+i\theta/2}\sin(\theta/2)\left|\phi_y\right\rangle$$

$$(4)$$

For the $E\otimes\varepsilon$ JT problem, **Fig. 1** depicts how $E(q_1,q_2)$ in the $q_2 = 0$ plane varies as a function of the distortion coordinate $q_1$. The potential energy surface, $E_\mp(q)$ vs. $q = |\bar{q}|$, has two branches; the lower branch has the shape of a *Mexican hat*, and the upper one that of a conical *Wizzard hat*.[11]

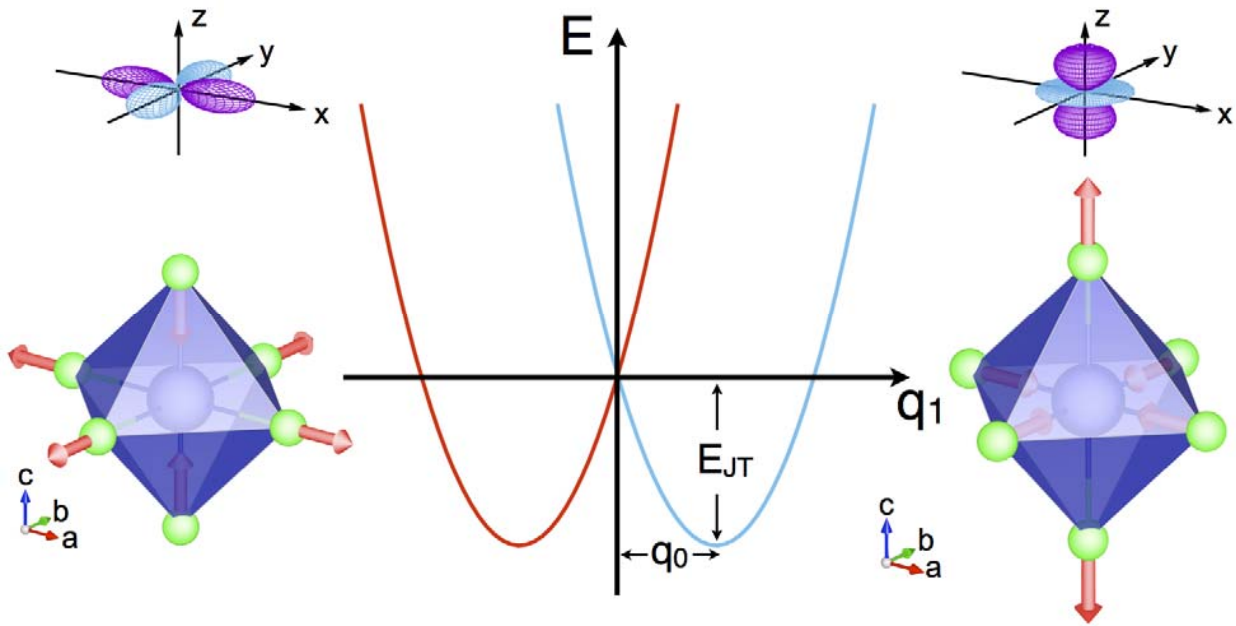

Figure 1. Section of the adiabatic potential energy surface for the $E\otimes\varepsilon$ JT problem in the $q_2 = 0$ plane. The character of the electronic wavefunction for positive and negative $q_1$

displacements is shown for the case of an octahedral complex with degenerate $\left|3z^2 - r^2\right\rangle$ and $\left|x^2 - y^2\right\rangle$ states. It has a conical intersection at $q_1 = q_2 = 0$.

It is important to note that, by employing the Pauli matrices $\sigma_i$ ($i$ = x, y, z) as well as $\sigma_0$ as the 2x2 unit matrix, the JT Hamiltonian $H_{JT}$ is rewritten as

$$H_{JT} = -\gamma(\sigma_z q_1 - \sigma_x q_2) + \frac{1}{2}Cq^2\sigma_0 \qquad (5)$$

Since the matrix Hamiltonian $H_{JT}$ is a linear combination of $\sigma_0$ and $\sigma_i$, $H_{JT}$ belongs to the group of special unitary matrices, SU(2), and the doublet ($\phi_x$, $\phi_y$) is a representation of the SU(2) group, i.e., a pseudo-spinor representation. For further discussion, see Supplementary Materials (SM):

$$\left|\phi_x\right\rangle \equiv \left|+\right\rangle = \begin{pmatrix} 1 \\ 0 \end{pmatrix}, \qquad \left|\phi_y\right\rangle \equiv \left|-\right\rangle = \begin{pmatrix} 0 \\ 1 \end{pmatrix} \qquad (6)$$

One can use $\sigma_z$ and $\sigma_x$ as the pseudo-spin operators in orbital space such that

$$\begin{aligned} \sigma_z\left|\phi_x\right\rangle = \left|\phi_x\right\rangle, \quad \sigma_z\left|\phi_y\right\rangle = -\left|\phi_y\right\rangle \\ \sigma_x\left|\phi_y\right\rangle = \left|\phi_x\right\rangle, \quad \sigma_x\left|\phi_x\right\rangle = +\left|\phi_y\right\rangle \end{aligned} \qquad (7)$$

with $\left|+\right\rangle$ and $\left|-\right\rangle$ as the eigenstates of $\sigma_z$ with eigenvalues $+1$ and $-1$, respectively. Then, upon changing the basis set, for instance, by introducing a new basis set

$$\left|\phi_1\right\rangle = \frac{1}{\sqrt{2}}\left(-\left|\phi_x\right\rangle + i\left|\phi_y\right\rangle\right), \qquad \left|\phi_2\right\rangle = \frac{1}{\sqrt{2}}\left(+\left|\phi_x\right\rangle + i\left|\phi_y\right\rangle\right) \qquad (8)$$

the JT Hamiltonian $H_{JT}$ is rewritten as

$$H_{JT} = \gamma \begin{pmatrix} 0 & q_1 - iq_2 \\ q_1 + iq_2 & 0 \end{pmatrix} + \frac{1}{2} Cq^2 \sigma_0 = \gamma(\sigma_x q_1 + \sigma_y q_2) + \frac{1}{2} Cq^2 \sigma_0$$

$$= \gamma(\vec{\sigma} \times \tilde{\vec{p}}) \cdot \hat{z} + \frac{1}{2} Cq^2 \sigma_0 \tag{9}$$

where the three Cartesian components of $\tilde{\vec{p}}$ are given by $(\tilde{p}_x, \tilde{p}_y, \tilde{p}_z) = (-q_2, q_1, 0)$. The mathematical structures describing the JT distortion are summarized in **Table 1**.

**Rashba spin splitting**

Let us now consider the Rashba spin splitting. For a nonmagnetic and centrosymmetric electronic system, the up-spin and down-spin states of a given energy band $\varepsilon(\vec{k}, \sigma)$ are degenerate. This is a consequence of time-reversal symmetry ($\Theta$), since the system is nonmagnetic and since the system has inversion symmetry ($\Pi$). Time-reversal implies that $\varepsilon(\vec{k}, \sigma) = \Theta \varepsilon(\vec{k}, \sigma) = \varepsilon(-\vec{k}, -\sigma)$. At $\vec{k} = 0$, namely, at $\Gamma$ point, $\varepsilon(0, \sigma) = \Theta \varepsilon(0, \sigma) = \varepsilon(0, -\sigma)$ so that the up-spin and down-spin states are degenerate at $\Gamma$. On the other hand, inversion symmetry implies $\varepsilon(\vec{k}, \sigma) = \Pi \varepsilon(\vec{k}, \sigma) = \varepsilon(-\vec{k}, \sigma)$. Then it follows that $\varepsilon(\vec{k}, \uparrow) = \varepsilon(\vec{k}, \downarrow)$, i.e., Kramers' degeneracy. The latter can be lifted if either time-reversal or inversion symmetry is broken. In the following we consider the case of breaking inversion symmetry.

For a simple two-dimensional (2D) free-electron gas represented by the electrons at a metallic surface or in the physics of low-dimensional semiconductor heterostructures,[12] the inversion symmetry is absent in the direction perpendicular to a 2D plane. (By convention, the z-axis is taken perpendicular to the surface.) With the introduction of a spin-orbit coupling (SOC) term, the dispersion relations of the resulting spin-split electronic states is essentially described by the Rashba Hamiltonian, $H_R$

$$H_R = \alpha_R(\vec{\sigma} \times \vec{k}) \cdot \hat{z} + \frac{\hbar^2}{2m^*}k^2\sigma_0 = \alpha_R(k_y\sigma_x - k_x\sigma_y) + \frac{\hbar^2}{2m^*}k^2\sigma_0 \qquad (10)$$

where $\alpha_R$ is the Rasbha coupling constant, $m^*$ the effective mass, and $\vec{k} = (k_x, k_y)$ the k-vector in reciprocal space. The essential physics is captured by this Hamiltonian, although the Rashba effect in solids can be derived within the framework of the $\vec{k} \cdot \vec{p}$ model or tight-binding formalism. These more sophisticated approaches show that atomic SOC is the essential ingredient leading to Eq. 10. In the $\vec{k} \cdot \vec{p}$ approach, consisting of a perturbative expansion of scalar relativistic Hamiltonian, the lowest term describes the coupling between $\vec{k}$ and the velocity operator $\vec{v} = 2\vec{p} + c^{-2}\vec{\sigma} \times \nabla V$ arising from the SOC term. In tight-binding models, the Rashba constant $\alpha_R$ is proportional to $t\Delta_{SO}$, where $t$ is the hopping integral between the orbitals of neighboring sites (see below for further discussion), and $\Delta_{SO}$ is the effective coupling constant resulting from the scalar-relativistic SOC, $\vec{L} \cdot \vec{S}$.

The Rashba Hamiltonian of Eq. 10 depends parametrically on the momentum $\vec{k} = (k_x, k_y) = (k\cos\theta, k\sin\theta)$, where $k = \sqrt{k_x^2 + k_y^2}$, and the doublet basis set $(\left|+\right\rangle, \left|-\right\rangle)$ represents the eigenstates of the $\sigma_z$ matrix, denoting spin-up and spin-down states with respect to a spin orientation axis perpendicular to the surface (hence parallel to the polar axis). By diagonalizing the $H_R$ one obtains the following eigenvalues

$$E_{\pm}(k) = \mp\alpha_R k + \frac{\hbar^2}{2m^*}k^2 \qquad (11)$$

where the second term describes a free-electron band dispersion and the first term the Rashba coupling with the material-dependent coefficient $\alpha_R$, which is taken to be positive without loss of generality. **Fig. 2a** shows the energy surface $E_\pm(k)$ as a function of the wave vectors $k_x = k\cos\theta$ and $k_y = k\sin\theta$, where the radius of the trough is given by $k_0 = \alpha_R m^* / \hbar^2$ with energy lowering $E_-(k_0) = -\alpha_R^2 m^* / 2\hbar^2$ so that the Rashba stabilization energy $E_R$ is $\alpha_R^2 m^* / 2\hbar^2$ (**Fig. 1**, **2a**). The eigenstates of $H_R$ can be written as

$$
\begin{aligned}
\left| \psi_- \right\rangle &= \frac{e^{i\vec{k}\cdot\vec{r}}}{2\pi} \frac{1}{\sqrt{2}} \left( \left| + \right\rangle + i e^{i\theta} \left| - \right\rangle \right) \\
\left| \psi_+ \right\rangle &= \frac{e^{i\vec{k}\cdot\vec{r}}}{2\pi} \frac{1}{\sqrt{2}} \left( \left| + \right\rangle - i e^{i\theta} \left| - \right\rangle \right)
\end{aligned}
\tag{12}
$$

An interesting consequence of a Rashba effect is the spin texture of the electronic states associated with the contour of k-points defined for a certain energy on the $E(k)$ vs. $k$ surface (Eq. 11). This is usually done by calculating the spin polarizations on the basis of the expectation values of the Pauli spin matrices, $\left\langle \psi_\pm(k) \left| \vec{\sigma} \right| \psi_\pm(k) \right\rangle$,

$$
\begin{aligned}
\left\langle \psi_\pm(k) \left| \sigma_x \right| \psi_\pm(k) \right\rangle &\equiv \left\langle \sigma_x \right\rangle_R = \pm \sin\theta(k) \\
\left\langle \psi_\pm(k) \left| \sigma_y \right| \psi_\pm(k) \right\rangle &\equiv \left\langle \sigma_y \right\rangle_R = \mp \cos\theta(k) \\
\left\langle \psi_\pm(k) \left| \sigma_z \right| \psi_\pm(k) \right\rangle &\equiv \left\langle \sigma_z \right\rangle_R = 0
\end{aligned}
\tag{13}
$$

As depicted in **Fig. 2a**, where the spin-expectation values are pictorially associated with three-dimensional arrows, the spin orientations for the Rashba Hamiltonian on one branch of the energy surface circle tangentially counterclockwise around $\Gamma$ in reciprocal space but those on the other branch circle tangentially clockwise. (The spin textures of **Fig. 2** are determined under the assumption that the coefficient $\alpha_R$ is positive.)

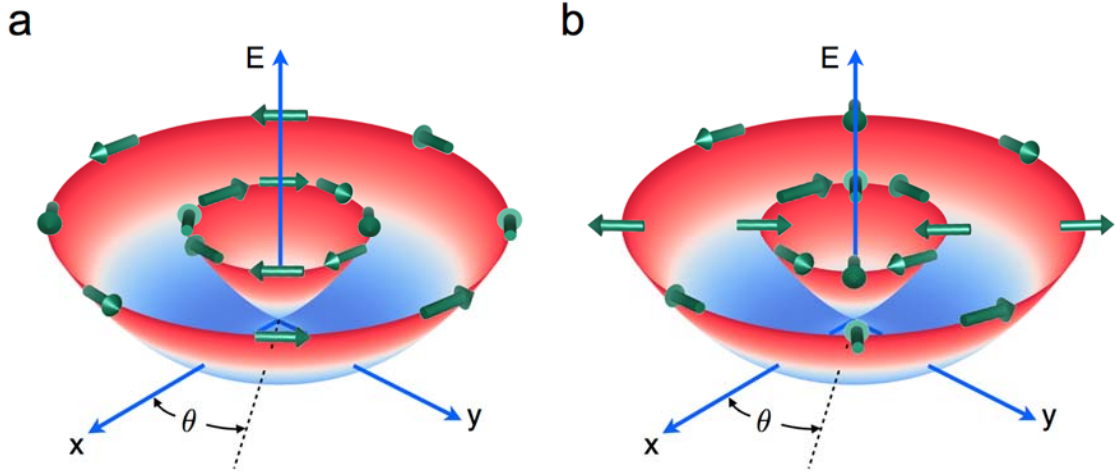

Fig. 2. Typical "Mexican-hat" shape of the adiabatic potential energy surface of the E⊗ε JT problem or that of the energy dispersion of electrons experiencing a Rashba interaction. The (x,y) axes label the $(q_1,q_2)$ and $(k_x,k_y)$ axes in the JT and Rashba systems, respectively, with θ as the corresponding polar angle. The arrows represent either the orbital texture or the spin texture: (a) tangential texture and (b) radial texture. To highlight the analogy between the JT orbital-polarization and the Rashba/Dresselhaus spin-polarization, $\langle \sigma_z \rangle_{JT}$ is plotted along the y-axis instead of the z-axis (cf. Eqs. 14 and 15 in the text).

**Comparison of the counterparts**

We now compare the conceptual and mathematical features of the JT and Rashba spin physics summarized in **Table 1**. The physics of both phenomena is represented by a 2×2 matrix Hamiltonian belonging to SU(2) so that the spin-formalism can be applied to both cases thereby leading to a doublet of orbital states and that of spin states. The variables describing the JT distortion are the normal modes $(q_1,q_2)$ defining the vibrations in real space, and those describing

Rashba effect are the wave vectors ($k_x$, $k_y$) in reciprocal space. $H_{JT}$ and $H_R$ share the same eigenvalues except for the rescaling of the constants; the coefficients $\hbar^2$ and $\alpha$ of Eq. 11 play the roles of the coefficients C and $\gamma$ of Eq. 3, respectively. In both cases, the dispersion relation in their space give rise to a Mexican hat for the lower branch and a Wizard hat for the upper branch, upon rotation by $2\pi$ around the energy axis. In both cases the conical intersection is retained only at the origin, $\vec{q} = 0$ and $\vec{k} = 0$, respectively (**Fig. 2**). Mathematically, the $H_{JT}$ and $H_R$ matrices can be related to each other by the similarity transformation using an appropriate unitary matrix.

Table 1. Mathematical structures of Jahn-Teller and Rashba physics

| | Jahn-Teller | Rashba |
|---|---|---|
| Variable | $\widetilde{p} = (-q_2, q_1)$ | $\vec{k} = (k_x, k_y)$ |
| Doublet | $\|\phi_x\rangle \equiv \|+\rangle = \begin{pmatrix} 1 \\ 0 \end{pmatrix}$, $\|\phi_y\rangle \equiv \|-\rangle = \begin{pmatrix} 0 \\ 1 \end{pmatrix}$ <br><br> Orbital doublet, pseudo-spin doublet | $\|+\rangle = \begin{pmatrix} 1 \\ 0 \end{pmatrix}$, $\|-\rangle = \begin{pmatrix} 0 \\ 1 \end{pmatrix}$ <br><br> Spin doublet |
| Hamiltonian | $H_{JT} = \gamma(\vec{\sigma} \times \widetilde{p}) \cdot \hat{z} + \frac{1}{2}Cq^2\sigma_0$ | $H_R = \alpha_R(\vec{\sigma} \times \vec{k}) \cdot \hat{z} + \frac{1}{2}\hbar^2 k^2\sigma_0$ |
| Eigenvalues | $E_{\mp}(q) = \mp\gamma q + \frac{1}{2}Cq^2$ | $E_{\mp}(k) = \mp\alpha_R k + \frac{1}{2}\hbar^2 k^2$ |
| Eigenstates | $\|\Psi_-\rangle = e^{-i\theta/2}\cos(\theta/2)\|\phi_x\rangle - e^{-i\theta/2}\sin(\theta/2)\|\phi_y\rangle$ <br><br> $\|\Psi_+\rangle = e^{+i\theta/2}\cos(\theta/2)\|\phi_x\rangle + e^{+i\theta/2}\sin(\theta/2)\|\phi_y\rangle$ | $\|\psi_-\rangle = \dfrac{e^{i\vec{k}\cdot\vec{r}}}{2\pi}\dfrac{1}{\sqrt{2}}\left(\|+\rangle + ie^{i\theta}\|-\rangle\right)$ <br><br> $\|\psi_+\rangle = \dfrac{e^{i\vec{k}\cdot\vec{r}}}{2\pi}\dfrac{1}{\sqrt{2}}\left(\|+\rangle - ie^{i\theta}\|-\rangle\right)$ |
| Polarization | $\langle\Psi_\pm(q)\|\sigma_x\|\Psi_\pm(q)\rangle \equiv \langle\sigma_x\rangle_{JT} = \pm\sin\theta(q)$ <br><br> $\langle\Psi_\pm(q)\|\sigma_y\|\Psi_\pm(q)\rangle \equiv \langle\sigma_y\rangle_{JT} = 0$ <br><br> $\langle\Psi_\pm(q)\|\sigma_z\|\Psi_\pm(q)\rangle \equiv \langle\sigma_z\rangle_{JT} = \mp\cos\theta(q)$ <br><br> Orbital polarization | $\langle\psi_\pm(k)\|\sigma_x\|\psi_\pm(k)\rangle \equiv \langle\sigma_x\rangle_R = \pm\sin\theta(k)$ <br><br> $\langle\psi_\pm(k)\|\sigma_y\|\psi_\pm(k)\rangle \equiv \langle\sigma_y\rangle_R = \mp\cos\theta(k)$ <br><br> $\langle\psi_\pm(k)\|\sigma_z\|\psi_\pm(k)\rangle \equiv \langle\sigma_z\rangle_R = 0$ <br><br> Spin polarization |

The conceptual features of the JT and Rashba physics are summarized in **Table 2**. The Rashba effect occurs when, for example, a 2D semiconductor loses inversion symmetry, due to the SOC of its constituent atoms. In a tight-binding description of the Rashba effect,[13] $\alpha_R \propto t\Delta_{SO}$, where $t$ is the hopping integral between the out-of-plane orbital $p_z$ of one atom with the in-plane orbitals $p_x/p_y$ of its adjacent atom. This hopping integral is nonzero in the absence of inversion symmetry. In the Rashba effect, degenerate states appear at $\Gamma$ because of time-reversal symmetry, which are split only away from $\Gamma$ due to the loss of inversion symmetry.

Table 2. Conceptual features of the Jahn-Teller and Rashba effects

| | Jahn-Teller | Rashba |
|---|---|---|
| Symmetry loss | Totally symmetric representation | Inversion symmetry |
| Required conditions | Unevenly filled degenerate electronic states | Spin-orbit coupling |
| Coupling | $\gamma = \langle \phi_1 \lvert \partial V / \partial q \rvert \phi_2 \rangle$ q-space orbital and vibration | $\alpha_R \propto t\Delta_{SO}$ k-space orbital and spin |
| Consequence | Split between degenerate electronic states | Split between up-spin and down-spin states |
| Coupled states | Orbital texture | Spin texture |
| Detection of texture | Time-resolved RIXS experiments? | Spin-resolved ARPES |

The JT distortion of a molecule occurs when its electronic structure is not totally symmetric with respect to its point symmetry group by having an unevenly-filled degenerate state. In such a case the vibronic coupling is not negligible, because the adiabatic approximation breaks down. Unevenly-filled degenerate electronic structures are not totally symmetric with respect to the molecular environment. Since they are coupled to nuclei via the vibronic interaction, the molecule experiences a symmetry-lowering distortion (for further discussion, see SM). In short, loss of inversion symmetry and SOC in the Rashba problem are equivalent to loss of totally symmetric representation and vibronic coupling in the JT effect.

As the counterpart of the Rashba spin texture, one can straightforwardly define the "orbital texture" in q-space of the JT electronic states. (for further discussion, see SM) and use it for an effective illustration of the pseudospin vectors. For the q-points on the contour defined for a constant energy of the E(q) vs. q surface (Eq. 3), one can calculate the orbital polarizations as the expectation values of the pseudo-spins introduced in Eqs. 6 and 7, namely,

$$\left\langle \Psi_{\pm}(q)\middle|\sigma_x\middle|\Psi_{\pm}(q)\right\rangle \equiv \left\langle \sigma_x\right\rangle_{JT} = \pm \sin\theta(q)$$
$$\left\langle \Psi_{\pm}(q)\middle|\sigma_y\middle|\Psi_{\pm}(q)\right\rangle \equiv \left\langle \sigma_y\right\rangle_{JT} = 0 \qquad (14)$$
$$\left\langle \Psi_{\pm}(q)\middle|\sigma_z\middle|\Psi_{\pm}(q)\right\rangle \equiv \left\langle \sigma_z\right\rangle_{JT} = \mp \cos\theta(q)$$

Note that $\left\langle \sigma_x\right\rangle_{JT}$ and $\left\langle \sigma_z\right\rangle_{JT}$ display exactly the same dependence on the polar angle $\theta$ as do $\left\langle \sigma_x\right\rangle_R$ and $\left\langle \sigma_y\right\rangle_R$ of the Rashba effect. The JT-split states appear as a mixing of the degenerate states, which depends parametrically on the JT distortions. As such, the orbital texture provides an effective graphical representation of JT-split states expressed in terms of the basis sets. The JT-induced orbital polarization effect eventually leads to orbital-ordering effects in many solid-state materials, for example, manganite oxides.[14,15] The cross-section of the JT potential energy surface at a certain energy below or above the degenerate point at $\vec{q} = 0$ consists of two concentric circles. Both circles belong to the Mexican hat below the degenerate point. Above the degenerate point, the outer circle belongs to the Mexican hat, but the inner one to the Wizard hat.

Thus, as the energy moves from below to above the degenerate point, the orbital texture of the inner circle switches its chirality from counterclockwise to clockwise (**Fig. 2a**). This maps one-to-one with the basics of Rashba physics: the spin texture of the inner Fermi surface (i.e., the inner cross-section of the Rashba energy surface) switches from counterclockwise below the degenerate point to clockwise above the degenerate point at $\vec{k} = 0$. The orbital textures determined for a model simulating an $E \otimes \varepsilon$ JT system on the basis of density functional calculations are presented in SM.

It is of importance to speculate how one might experimentally detect the orbital texture associated with the JT effect. Numerous studies have been concerned with straightforward observations of Jahn-Teller distortions.[5] However, detecting the orbital nature of the ground and excited states of a JT system becomes more subtle, because it is generally linked to a more complicated Hamiltonian and to the possible occurrence of cooperative effects and orbital ordering.[16,17] Here we address a fundamentally different issue; can we energy-scan the orbital-phonon entanglement of the excited states of a JT center? The orbital texture is a manifestation of the entanglement between the orbital and phononic degrees of freedom, so its detection requires a simultaneous access to the local phonon and orbitally resolved electronic spectra. Thus, it would be interesting to perform a typical pump-probe experiment, in which one might excite the JT-phonons and probe the orbitally resolved electronic states or, conversely, excite the orbital degrees of freedom and probe the JT-modes. The resonant inelastic x-ray scattering (RIXS) spectroscopy [16] is extremely sensitivity to pure electronic and phonon excitations. To probe the $E \otimes \varepsilon$ vibronic coupling, for example, in a JT-active $MO_6$ octahedron (M = transition-metal element), one might perform time-resolved RIXS by pumping at the K-edge of the oxygen (1s→2p) to excite the JT phonons of $MO_6$ and then analyzing the relaxation of the orbital polarization. The latter can be done by resolving the electric polarization of the L-edge of the transition metal (2p→3d). According to the energy of the excited phonons in terms of the energy of the pump (accessing states above the Mexican-hat minimum or the Wizzard-hat minimum), it should be possible to see a different relaxation of the net orbital polarization. Then, the scanning the energy window between the Mexican-hat and Wizzard-hat minima compared with that above the Wizzard-hat minimum can provide information about the change in the orbital texture (for further discussion, see SM).

**Generalization**

So far we focused on the analogies between two specific problems, namely, the E⊗ε JT problem and the Rashba problem for a 2D electron gas. In general, Rashba-like effects may appear at the surface of solid-state materials containing heavy elements with large SOC, where the explicit form of the spin-momentum coupling is dictated by the crystalline symmetries of the periodic lattice (in fact, Rashba-like spin splitting and spin polarizations can also be realized at high-symmetry k-points where time-reversal symmetry is broken).[17] As a practical example, let us consider the linear Dresselhaus effect. The latter, similar to a Rashba effect, arises from the loss of a reflection symmetry about a plane containing at least one lattice site of the underlying crystal. Specifically, the associated energy surface is given by exactly the same expression, Eq. 12 (depicted in **Fig. 2**), whereas the spin texture associated with a Dresselhaus effect is of radial type (**Fig. 2b**) around a certain k-point in contrast to the tangential one (**Fig. 2a**) found for a Rashba effect. The coupling term typically describing a linear Dresselhaus effect is given by

$$H_D = \alpha_D(-k_x\sigma_x + k_y\sigma_y) \tag{16}$$

where $\alpha_D$ is the coupling constant that is system-dependent. When Rashba and Dresselhaus effects are both present, with one coupling stronger than the other (a situation generally met at k-points with $C_{2v}$ symmetry),[18] the dispersion relations of the resulting spin-split states become asymmetric and look as if they were generated by two parabolic cones overlapping partially (see **Fig. 3a**),[9] with the coupling Hamiltonian given by

$$H_{RD} = \alpha_R (k_y \sigma_x - k_x \sigma_y) + \alpha_D (-k_x \sigma_x + k_y \sigma_y) \qquad (17)$$

Dresselhaus-like spin textures (**Fig. 2b**), which are realized for linear Dresselhaus coupling, have their analogues in the $E \otimes \varepsilon$ JT problem in molecular systems with $D_5$ and $D_6$ point group symmetries (**Fig. 2b**). For the JT vibronic interaction given by Eq. 5 is modified as $\gamma(\sigma_z q_1 + \sigma_x q_2)$, so that the expectation values of the orbital pseudospin become $\langle \sigma_z \rangle_{JT} = \pm \cos \theta(q)$ and $\langle \sigma_x \rangle_{JT} = \pm \sin \theta(q)$. In the JT distortions the explicit form of the vibronic coupling is dictated by the specific point-group symmetries of the molecules.[5] It may happen that a degenerate orbital doublet interacts with two nondegenerate phonon modes (e.g., $q_1$ and $q_2$ are not degenerate in **Fig. 2**), which is labeled as an $E \otimes (b_1 + b_2)$ JT problem (**Fig. 3**).[9]

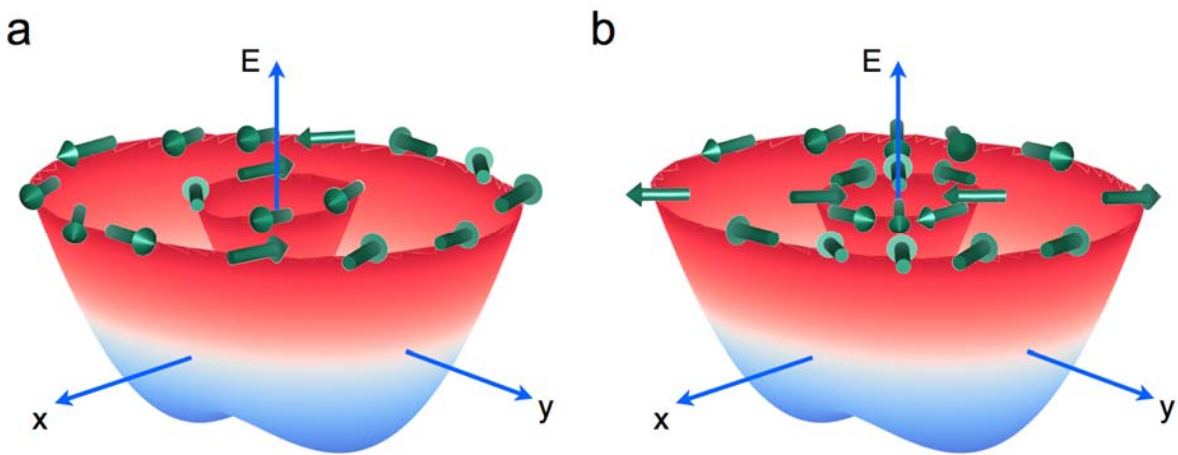

Fig. 3. Energy surfaces for the Rashba-Dresselhaus model (realized, e.g., around k-points with $C_{2v}$ symmetry) and the $E \otimes (b_1+b_2)$ JT problem (as found, e.g., for molecular complexes with $C_{4v}$ or $D_{2d}$ point-group symmetries). The spin and orbital textures show an analogous behavior, with (a) a more tangential-like and (b) a more radial-like texture. The tangential or radial character of the textures depends on the relative sign of the two competing interactions (namely, Rashba vs. Dresselhausin spin physics, and $b_1$ vs. $b_2$ interactions in JT physics). For instance, the JT vibronic interaction is described by $\pm \gamma_1 q_1 \sigma_z + \gamma_2 q_2 \sigma_x$, with the $\pm$ sign found, e.g., for $D_{2d}$ and $C_{4v}$, respectively.

So far our discussion of the analogies considered only the linear coupling term in both effects. Higher-order contributions in Rashba-like problems always display odd powers of the momentum k, whereas quadratic contributions are typically considered in the JT problems. The potential energy surface of the $E \otimes \varepsilon$ JT problem is warped when the quadratic vibronic coupling, $\langle \phi_1 | \partial^2 V / \partial q_1 \partial q_2 | \phi_2 \rangle$, is included, leading to the formation of three minima along the bottom of the trough of **Fig. 2**.[7] Likewise, the inclusion of cubic coupling term in the Rashba Hamiltonian generally leads to a hexagonal warping with six minima appearing in the band dispersion shown in **Fig. 2**.[19]

Finally, it is of interest to consider the coupling interactions of the JT and Rashba effects from the viewpoint of fermionic and bosonic character. The vibronic coupling taking place in the JT effect is by definition a fermion-boson interaction between electronic and phononic degrees of freedom, whereas the Rashba SOC has an intrinsic fermionic nature. However, it has been recently realized that Rashba effects in confined systems, such as quantum dots and ultracold atomic gases in a trapping magnetic or optical potential may be described by spin-boson models. The latter are formally equivalent to JT linear problems and describe the interaction between a bosonic mode (e.g., the trapping potential) and a fermionic two-level system.[20]

**Conclusions**

In summary, we discussed a formal mathematical analogy between the physics of JT distortion and that of Rashba spin splitting. Our exploration of the conceptual features between the two physics allowed us to discuss into the JT theory the orbital texture, which is the counterpart of the spin texture in Rashba spin physics. As a possible way of experimentally detecting the orbital texture, we propose time-resolved RIXS experiments. It is hoped that this work will stimulate further discussions between the two physics communities.

**Acknowledgments**

M.H.W. would like to thank Dr. Silvia Picozzi for her kind invitations to visit her SPIN-CNR laboratory in Chieti and L'Aquila, and acknowledges support by the CNR Short-Term Mobility Program 2015. A.S. would like to thank Prof. C. Franchini and G. Kresse for their kind invitation to University of Vienna. D.D.S., S.P. and A.S. acknowledge the CARIPLO Foundation through the MAGISTER project Rif. 2013-0726. The figures were done by D.D.S. using the VESTA[21] and Gnuplot[22] programs.